# Cluster Model for parsimonious selection of variables and enhancing Students' Employability Prediction

Pooja Thakar, Research Scholar

Department of Computer Science
Banasthali University
Jaipur, Rajasthan, India

Prof. Dr. Anil Mehta, Professor

Department of Management
University of Rajasthan
Jaipur, Rajasthan, India

Dr. Manisha, Associate Professor

Department of Computer Science
Banasthali University
Jaipur, Rajasthan, India

*Abstract*—**Educational Data Mining (EDM) is a promising field, where data mining is widely used for predicting students' performance. One of the most prevalent and recent challenge that higher education faces today is making students skillfully employable. Institutions possess large volume of data; still they are unable to reveal knowledge and guide their students. Data in education is generally very large, multidimensional and unbalanced in nature. Process of extracting knowledge from such data has its own set of problems and is a very complicated task.**

**In this paper, Engineering and MCA (Masters in Computer Applications) students data is collected from various universities and institutes pan India. The dataset is large, unbalanced and multidimensional in nature. A cluster based model is presented in this paper, which, when applied at pre-processing stage helps in parsimonious selection of variables and improves the performance of predictive algorithms. Hence, facilitate in better prediction of Students' Employability.**

## I. INTRODUCTION

In recent times, Education to a greater extent has become employment oriented. Placements of an institution build its reputation and market value. Employability of students' not only plays a very vital role in their own growth, but is a crucial factor for the growth of institute and nation by enlarge.

Educational Data Mining is a field that develops methods to discover knowledge from students' data. Institutes want to identify those students, who have the potential to fetch best placement in the market. They also want to know the attributes of best achievers, so that others can be guided and trained to perform better. Predicting student's employability in the beginning stage, as soon as they enter the institute can be used to take proactive and timely actions. This may improve their performance and will also identify students, who are at the risk of unemployment.

Institutes manage colossal records of students. Along with basic primary academic data; institutes also maintain secondary cognitive, psychometric and analytical records of students. Factors such as communication skills, personal qualities, teamwork, critical thinking, and problem solving skills are important employability factors [1]. Studies also reflect the importance of secondary attributes along with primary attributes for higher prediction accuracy and analysis [23]. Factors like location, medium of teaching, qualification of mother, income of family and status are highly correlated with academic performance of students [25]. Such a data has high potential to uncover knowledge that can be utilized to guide students for better employability. But, the data collected from educational ecosystem suffer with two intrinsic problems. They are:

a) Data is Unbalanced – High Achievers records are very less as compared to Non Achievers, thus unbalanced, which is further difficult to predict and extract rules. Data is unbalanced, when some classes have significantly large number of instances available than others. Predictive algorithms overlook minority classes, and considers only majority classes. Minority class hubs may lead to misclassification in many high-dimensional datasets [3]. As a result, the predictive algorithms are unable to classify them correctly.

b) Data is Multidimensional: With inclusion of Personal, Geographical, Family Background, Professional, Academic , Psycometric, Cognitive, Non Cognitive attributes of every student, data become multidimensional. Controlled dataset with high dimensionality may result in over-fitting and decrease the generalization performance of predictive algorithms[2]. This in turn requires huge set of data for better prediction.

Teaching –Learning Systems are also highly customized to fulfill the needs of specific Institute for which it is designed. There is no unified model that can be used across institutes with any type of dataset.

In Indian higher education system, Engineering (B.E. / B.Tech.) and MCA (Masters in Computer Applications) are two professional degree courses that makes students ready for Information Technology (IT) based Companies. Most of the students in these courses aspire and compete to obtain the best campus placements, before they pass out.

This paper presents a model that uses clustering at pre-processing stage and improves the performance of various predictive algorithms to predict employability. Data is collected from various universities and institutes pan India by contacting Training and Placement Cell of the Institutes. Dataset is huge with the record set of around 8973 students, multidimensional with 152 attributes and unbalanced due to very less records of students placed in top companies. RapidMiner Studio Academia Version 7.1.001 is used to apply various data mining algorithms.

Predictive Algorithms when first applied on raw dataset didn't produce good prediction results. Then clustering model is applied at pre-processing stage. It helps in reducing dimensionality and increasing prediction performance. Comparative results show significant differences and clearly demonstrate, how clustering when applied at pre-processing stage boosts the performance of predictive algorithms.

Further, this paper is structured as follows: Section II presents review of literature Section III describes data mining, clustering and classification techniques used in the study Section IV presents experimental setup, Section V demonstrates the results with comparative analysis and charts. Section VI concludes with scope for future work.

## II. REVIEW OF LITERATURE

In last decade, Data Mining has aroused the research interest of researchers in the field of Education and majority of them have worked to determine the techniques in predicting academic performance [15] [16] [17] [18] [19] [33][34].

In last few years, researchers have tried to find out the factors affecting employability of students. Emotional Intelligence, Personal Management, Life experience, Work Experience are considered as important factors for Employability Development Profile [5]. Emerald Group Publishing Limited published a paper in 2014 that described link between Competences, Dispositions and Employability [6]. Cairns, Gueni, Fhima, David and Khelifa found positive correlation in employees jobs/assignments, history etc. [7]. Potgieter & Coetzee revealed significant relationships between the personality and employability. They listed eight core employability features for fetching and sustaining employment opportunities [8]. Employers give premier importance to soft-skills and the lowest weight on academic reputation as described by David, Hamilton, Riley and Mark [9]. Denise Jackson and Elaine Chapman indicated significant differences between academic and employer skill ratings and revealed major skill gap between academia and industry [10]. V. K. Gokuladas necessitated specific skills in students beyond general academics, he showcased that GPA and proficiency in English language are vital predictors of employability [11] [12]. Noor Aieda, Abu Bakar, Aida Mustapha, Kamariah Md. Nasir found that graduates by enlarge lack in interpersonal communication, creative and critical thinking, problem solving, analytical, and team work [13]. Jamaludin, Nor Azliana, and Shamsul Sahibuddin highlighted that industry requires skilled worker to handle the projects as compared to only CGPA achievement [32].

Recently, few researchers have worked on students' employability prediction. In 2016, Mishra, Kumar and Gupta collected the data of 1400 MCA students to predict students' employability in campus placements excluding cognitive factors such as reasoning, aptitude and communication skill [20]. Piad, K.C., Dumlao, M., Ballera, M.A. and Ambat, S.C. published a paper in 2016 to predict employability of Information Technology students. Dataset of 515 students with 9 variables was taken for prediction [21]. Awang-Hashim, R., Lim, H.E., Yatim, B., TENGKU ARIFFIN, T.F., Zubairi, A.M., Yon, H. and Osman, O. in 2015 designed a statistical prediction model to identify low employability graduates in Malaysia [22]. In a recent study of 2016, conducted by Chen, Pei-lin, Wei Cheng, and Ting-ting Fan collected the dataset of 2160 with 6 attributes such as gender, area, type of college, degree, master and grades were used to create an employability model [24]. Jantawan, Bangsuk, and Cheng-Fa Tsai, designed a model for employability prediction using data mining techniques for a university in Thiland and emphasized on the need of more variables and automated preprocessing, [14].

Most of the researches emphasized to include all types of parameters such as cognitive, psychometric, background with academic attributes for better prediction [1][8][9][11][14][23][25]. Need for large, multidimensional dataset with automated pre-processing is also reflected [14].

Recent study by Tarik A. Rashid in 2015 pointed out future works to be based upon three parameters, first is to increase the size and type of the dataset, finding different ways to work out the problem with unbalanced data and examine other feature selection techniques that could enhance the performance of classification techniques [35].

This paper is an attempt to deal with the problem of large, multidimensional, unbalanced dataset of education system. Automated and speedy feature selection at preprocessing is performed to enhance the performance of predictive techniques.

## III. DATA MINING, CLUSTERING AND CLASSIFICATION TECHNIQUES

As quoted by Han and Kamber "Data mining refers to extracting or mining knowledge from large amounts of data." [27]. There are two methods in data mining for learning and extracting knowledge. They are supervised and unsupervised learning. Supervised learning or commonly known as classification is based on supervision i.e. the training data is accompanied by class and new data is classified on the basis of training set [26]. In unsupervised learning or commonly known as clustering, the class labels of training data is unknown [26].

The Model presented in this paper uses clustering at preprocessing stage to enhance the performance of classification. The techniques used in model are described below:

k-Nearest Neighbor: It is better known as learning by analogy and is the simplest of all. It compares a given test example with training examples for similarity. All of the training instances are stored in an n-dimensional pattern space. For new example, a k-nearest neighbor algorithm searches the pattern space for k training examples that are closest to the new example. These k training examples are the k "nearest neighbors" of the new example. Distance metric, such as the Euclidean distance is used [4]. The basic k-Nearest Neighbor algorithm is composed of two steps: Find the k training examples that are closest to the new example. Take the most commonly occurring classification for these k examples [4].

Naive Bayes (Kernel): Naïve Bayes classifier is also known as probabilistic classifier. It applies Bayes' theorem with strong independence assumptions. It assumes the presence or absence of a particular attribute of a class is unrelated to the presence or absence of any other attribute. Naive Bayes classifier considers all the properties to independently contribute to the probability. Kernels are used in kernel density estimation to estimate random variables density functions [4].

Decision Tree: A decision tree is more like an inverted tree. The representation is more meaningful and easy to interpret. It creates a classification model that predicts the value of a target attribute. Decision Trees are generated by recursive partitioning. The recursion stops when all the examples or instances have the same label value, i.e. the subset is pure or recursion may stop, if most of the examples are of the same label value. Pruning is a technique in which leaf nodes that do not add to the discriminative power of the decision tree are removed [4].

Neural Nets: The perceptron is a type of artificial neural network, simplest kind of feed-forward neural network; a linear classifier. For all incorrectly classified data points, the weight vector is either increased or decreased by the corresponding example values [4].

An artificial neural network (ANN), usually called neural network (NN), is a mathematical model inspired by biological neural networks. A neural network is composed of an interconnected group of artificial neurons. It processes information using a connectionist approach to computation. ANN is an adaptive system with changing internal structure based on information (internal or external) that flows through the network during the learning phase. A feed-forward neural network is one, where connections among units do not form a directed cycle. In this network, the information moves in only one direction, forward, from the input nodes, through the hidden nodes to the output nodes [4].

Perceptron learns a model by means of a feed-forward neural network trained by a back propagation algorithm. (multi-layer perceptron). Back propagation algorithm works in two phases: propagation and weight update. The two phases are repeated until the performance of the network is good enough. Algorithm adjusts the weights of each link in order to reduce the value of the error function by some small amount.

Network will finally converge to some state where the error of calculation is small. A multilayer perceptron (MLP) consists of multiple layers of nodes in a directed graph, with each layer fully connected to the next one. Each node is a neuron with a nonlinear activation function except inner nodes [4]. In this usual sigmoid function is used as the activation function. Therefore, the values ranges of the attributes should be scaled to -1 and +1[4].

Support Vector Machine: SVM is a powerful method for both classification and regression. The libsvm supports internal multiclass learning and probability estimation based on Platt scaling for proper confidence values. A SVM constructs a hyperplane or set of hyperplanes in a high/infinite dimensional space, which can be used for classification. A good separation is achieved by the hyperplane intuitively that has the major distance to the nearest training data points of any class, since in general the larger the margin the lower the generalization error of the classifier [4].

Linear Discriminant Analysis: LDA tries to find the linear combination of features, which best separates two or more classes of examples. The resulting combination is then used as a linear classifier. LDA explicitly attempts to model the difference between the classes of data. Discriminant analysis may be used for two objectives: either we want to review the adequacy of classification; or we assign objects to one of known groups of objects. Discriminant Analysis may thus have a descriptive or a predictive objective [4].

## IV. EXPERIMENTAL SETUP

### A. Data Collection

Data set used in this study is obtained from various Universities and Institutes pan India. Training and Placement Cells of various Institutes offering four year Engineering (Bachelor in Engineering – B.E. or Bachelor in Technology - B.Tech.) course or three years MCA (Master in Computer Applications) Degree Course were contacted. TPO (Training and Placement Officers) sent their institutes past records in the format provided to them in spreadsheet. Many of them provided data in the format they keep it at the Institute level. Placed Students details were either provided by the TPO or secondary data is collected from the Institutes website. Data collected is then compiled in one spread sheet.

### B. Data Selection and Transformation

Data is deliberately collected for Engineering and MCA students keeping in mind that maximum of them opt for placements after obtaining degree and in both courses last semester is specifically dedicated to internship in Companies. This is to ensure seriousness in students, while they sit for campus placements. Graduate courses such as BCA (Bachelor in Computer Applications) are not included as students are not very sincere for the placements; most of them opt for higher studies after obtaining degree. The dataset thus obtained comprises of around 8973 instances with 152 attributes. Dataset is large, unbalanced and multidimensional in nature.

Before proceeding for mining, one level pre-processing is done by excluding not pertinent attributes such as name of student, name of university/institute, course, batch, phone number. Derived variables were obtained like age. Most of the attributes were made consequential by altering them into categories like addresses converted to states and marks converted to grades. Dataset is later converted to its numerical counterpart for efficient clustering. It is named as "RD". Predictors includes Personal Predictors; such as Gender, Age, State, Family Background; such as Type of Family, Number of Siblings, Mother's Qualification/Occupation, Father's Qualification/Occupation , Academic Predictors; such as Percentage in secondary, senior secondary, graduation, gap in studies, Number of supplementary subjects, Aptitude Predictors; such as Command in English, Quantitative, Reasoning, Psychometric Predictors; such as conscientiousness, agreeableness, extraversion, neuroticism have been taken for the study.

### C. Implementation of Clustering and Classification Algorithms

RapidMiner Studio Academia Version 7.1.001 is used to implement machine leaning algorithms. k-means (Fast) clustering algorithm is used at the stage of pre-processing with numerical measure of Jaccard Similarity. k-means algorithm with Jaccard distance measure improves the clustering performance [31].

The algorithms used for prediction on the present dataset are Lazy- k-NN, Bayesian - Naïve Bayes Kernel, Neural Nets – Perceptron, AutoMLP, Neural Net, Support Vector Machine –LibSVM, HyperHyper, Fast Large Margin, Trees – Decision Tree, Logistic Regression, Discriminant Analysis - Linear (LDA).

The 10-fold cross-validation is chosen as an estimation approach to obtain a reasonable idea of classifier performance, since there's no separate test data set. This technique divide training set into 10 equal parts, 9 are applied as training set for making machine algorithm learn and 1 part is used as test set. This approach is enforced 10 times on same dataset, where every training set act as test set once.

### D. Performance Measures

Classification accuracy is the number of correct predictions made, divided by the total number of predictions made, multiplied by 100 to turn it into a percentage. In a problem where there is a large class imbalance, a model can predict the value of the majority class for all predictions and achieve high classification accuracy. This is called the Accuracy Paradox [28]. The dataset used in study suffers with this; hence accuracy measure may not be the perfect indicator to judge the performance of classifiers. Classification accuracy is typically not enough information to make decision on effectiveness of model if dataset has unbalanced classes.

F1 Score is the weighted average of Precision and Recall. Therefore, this score takes both false positives and false negatives into account. F1 is more useful than accuracy, if you have an uneven class distribution [29].

The formula for F1 Score is 2*(Recall * Precision)/(Recall + Precision).

Kappa is also another measure, which may be used in this case. Kappa Statistics is a normalized value of agreement for chance. It can be described as

$$K= (P (A)- P(E))/(1-P (E))$$

Where,

P (A) is percentage agreement and P (E) is chance agreement.

If K =1 than agreement is ideal between the classifier and ground truth.

If K=0, it indicates there's a chance of agreement. [30]

### E. Cluster Model for pre-processing

A novel approach is adopted for pre-processing the raw data (RD). Dataset is first transposed by swapping attributes as instances and instances as attributes. The new dataset now obtained is named as (RD1). K-means Clustering with Jaccard Similarity is applied on raw dataset (RD1), which results in reducing dimensionality, find relevant set of attributes very fast and improves classification performance.

Step wise description of model is as follows:

Step1: Retrieve Raw Dataset (RD) in the form of matrix (Raw dataset after cleaning and necessary transformations)

Step2: Transpose (TP) matrix RD i.e. RD1=TP [RD]

Step3: Apply clustering on RD1 with k = 2 or 3 (Apply k-means-Fast clustering algorithm with Jaccard Similariry on RD1)

Step4: Perform Filtering of RD1 w.r.t. Clusters

Step 5: Generate new Dataset w.r.t. Clusters named as Data Cluster (DC1), DC2 and DC3

Step6: Transpose all Data Clusters as Cluster1=TP [DC1] Cluster2=TP [DC2] Cluster3=TP [DC3]

Step7: Implement selected Classification Algorithm on each data cluster Cluster1, Cluster2 and Cluster3

Step8: Validate Performance of classifier with X-Validation by applying 10 fold cross validation.

## V. RESULTS

The performances of eleven predictive algorithms for predicting students' employability on the aforesaid dataset were experimented upon and results were calculated. Due to intrinsic problems of large, unbalanced, multidimensional dataset the results obtained were not satisfactory, when only predictive algorithms were applied directly on Raw Data (RD).

To overcome the problem of multidimensional data, PCA (Principle Component Analysis) is considered as very versatile, oldest and the most popular technique in multivariate analysis [36]. But when applied with the aforementioned dataset, it could not improve the results much. Comparative is shown in Table I and Table II below.

Cluster Model is then applied on this large, multidimensional unbalanced dataset; k-means Clustering with Jaccard s imilarity is applied on the set of 152 attributes (after transposing the data) at the level of pre-processing. This divided the dataset into clusters. The clusters now obtained contain reduced set of attributes, which are correlated to each other. This method eventually reduced the attribute set at a very fast pace. Now each cluster is transposed again and clustering algorithm is applied on them. Validated the performance of classifier with 10-fold cross validation and it shows significant improvement in results. It is also noticed that the cluster, which has Class as an attribute shows best results. Comparative results are as follows:

TABLE I: COMPARATIVE TABLE OF F1 SCORE

| | | F1 Score | | |
|---|---|---|---|---|
| S.No | Predictive Algorithm | Only Predictive Algorithm | Predictive Algorithm with PCA | Cluster Model |
| 1 | k-NN | 26.9 | 44.6 | 44.7 |
| 2 | Naïve Bayes Kernel | 55.1 | 41.4 | 50.1 |
| 3 | Perceptron | 39.6 | 32.1 | 42.5 |
| 4 | AutoMLP | 39.8 | 39.2 | 49.8 |
| 5 | Neural Net | 48.7 | 43.25 | 51.2 |
| 6 | LibSVM | 24.5 | 23.1 | 33.7 |
| 7 | Hyper Hyper | 37.2 | 37.3 | 38.5 |
| 8 | Fast Large Margin | 37.3 | 32.1 | 34.9 |
| 9 | Logistic Regression | 23.7 | 6.3 | 36.3 |
| 10 | LDA | 0 | 21.2 | 30 |
| 11 | Decision Tree | 0 | 2 | 1.9 |

Table I and Fig. 1 of F1 Score evidently exhibit the improvement in F1 score of each predictive algorithm, when Cluster Model is applied. PCA when applied with predictive algorithm improves results in some cases, but clearly performs less as compared to Cluster Model.

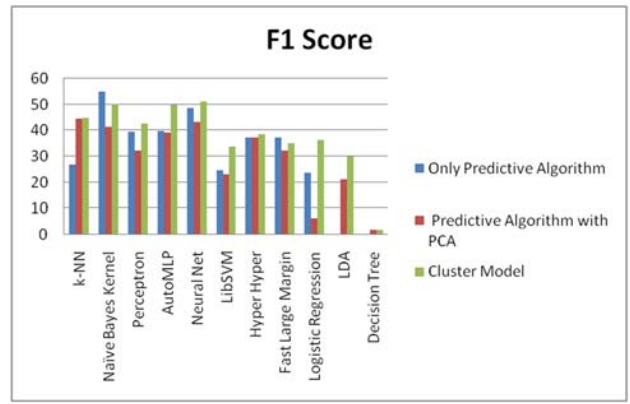

Fig. 1. Graph depicting Classifiers Performance with F1

This can be further proved with Statistical t-test measures. Results reflect that the improvement is significant.

Let's define hypothesis:

Hypothesis 1.1
Ho 1.1: There is no significant difference in F1 score of various Predictive Algorithms
$H_A$ 1.1: There is significant difference in F1 Score of various Predictive Algorithms.

Hypothesis 1.2
Ho 1.2: There is no significant difference in Kappa values of various Predictive Algorithms
$H_A$ 1.2: There is significant difference in Kappa values of various Predictive Algorithms.

t-Test: Paired Two Sample for Means is conducted between the F1 Score obtained by Predictive Algorithms only and when the same algorithms were applied using cluster model. Results are as follows

t-Test: Paired Two Sample for Means

| | Only Predictive Algorithm | Cluster Model |
|---|---|---|
| Mean | 30.25454545 | 37.6 |
| Variance | 315.1867273 | 192.172 |
| Observations | 11 | 11 |
| Pearson Correlation | 0.824050686 | |
| Hypothesized Mean Difference | 0 | |
| df | 10 | |
| t Stat | -2.415233157 | |
| P(T<=t) one-tail | 0.018178345 | |
| t Critical one-tail | 1.812461102 | |
| P(T<=t) two-tail | 0.036356689 | |
| t Critical two-tail | 2.228138842 | |

Results depict that t Stat > t Critical two-tail.

At 5% level of significance Ho 1.1 is rejected; hence there is a significant difference between F1 scores of predictive Algorithms when applied directly as compared to same predictive algorithms when applied using Cluster Model. Cluster Model outperforms.

There is also significant difference between variance values. Variance reduces to a greater extent.

Further, well known PCA method for dimensionality reduction is also compared with Cluster Model for its performance. Results show that Cluster Model performs better than PCA when applied on predictive algorithms.

t-Test: Paired Two Sample for Means is further conducted between the results obtained by Predictive Algorithms with PCA and when the same algorithms were applied with cluster model. Results are as follows

t-Test: Paired Two Sample for Means

|  | Predictive Algorithm with PCA | Cluster Model |
|---|---|---|
| Mean | 29.32272727 | 37.6 |
| Variance | 213.1126818 | 192.172 |
| Observations | 11 | 11 |
| Pearson Correlation | 0.827185965 |  |
| Hypothesized Mean Difference | 0 |  |
| df | 10 |  |
| t Stat | -3.269866894 |  |
| P(T<=t) one-tail | 0.0042161 |  |
| t Critical one-tail | 1.812461102 |  |
| P(T<=t) two-tail | 0.008432199 |  |
| t Critical two-tail | 2.228138842 |  |

Results depict that t Stat > t Critical two-tail

At 5% level of significance Ho 1.1 is rejected; hence there is a significant difference between F1 scores of predictive Algorithms when applied using PCA at preprocessing as compared to same predictive algorithms when applied using Cluster Model at preprocessing. Cluster Model transcends.

There is also significant difference between variance values. Variance decreases to a greater extent.

To clarify the results further, comparisons between Predictive only and Predictive Algorithms with PCA is also conducted. The results depict that PCA could not perform better on such dataset, which is large, multidimensional and unbalanced in nature.

t-Test: Paired Two Sample for Means is conducted between the results obtained by Predictive Algorithms only and when the same algorithms were applied with PCA. Results are as follows

t-Test: Paired Two Sample for Means

|  | Only Predictive Algorithm | Predictive Algorithm with PCA |
|---|---|---|
| Mean | 30.25454545 | 29.32272727 |
| Variance | 315.1867273 | 213.1126818 |
| Observations | 11 | 11 |
| Pearson Correlation | 0.756261858 |  |
| Hypothesized Mean Difference | 0 |  |
| df | 10 |  |
| t Stat | 0.264720359 |  |
| P(T<=t) one-tail | 0.398300577 |  |
| t Critical one-tail | 1.812461102 |  |
| P(T<=t) two-tail | 0.796601154 |  |
| t Critical two-tail | 2.228138842 |  |

Results depict that t Stat < t Critical two-tail

At 5% level of significance Ho 1.1 is accepted; hence there is no significant difference between F1 scores of predictive Algorithms when applied directly as compared to same predictive algorithms when applied using PCA There is also not much difference between variance values.

All the above results prove that Cluster Model when applied at preprocessing stage with predictive algorithms improves F1 Score significantly. Hence enhances the performance of predictive algorithms to predict students' employability even with complex dataset. This can further be illustrated with Kappa values.

TABLE II: COMPARITIVE TABLE OF KAPPA SCORE

| S.No | Predictive Algorithm | Kappa | | |
|---|---|---|---|---|
|  |  | Only Predictive Algorithm | Predictive Algorithm with PCA | Cluster Model |
| 1 | k-NN | 0.03 | 0.242 | 0.249 |
| 2 | Naïve Bayes Kernel | 0.364 | 0.214 | 0.281 |
| 3 | Perceptron | 0.19 | 0.049 | 0.171 |
| 4 | AutoMLP | 0.282 | 0.264 | 0.368 |
| 5 | Neural Net | 0.348 | 0.292 | 0.365 |
| 6 | LibSVM | 0.155 | 0.136 | 0.228 |
| 7 | Hyper Hyper | 0.018 | 0.021 | -0.003 |
| 8 | Fast Large Margin | 0.149 | 0.067 | 0.132 |
| 9 | Logistic Regression | 0.137 | 0.042 | 0.232 |
| 10 | LDA | 0 | 0.135 | 0.199 |
| 11 | Decision Tree | 0 | 0.014 | 0.013 |

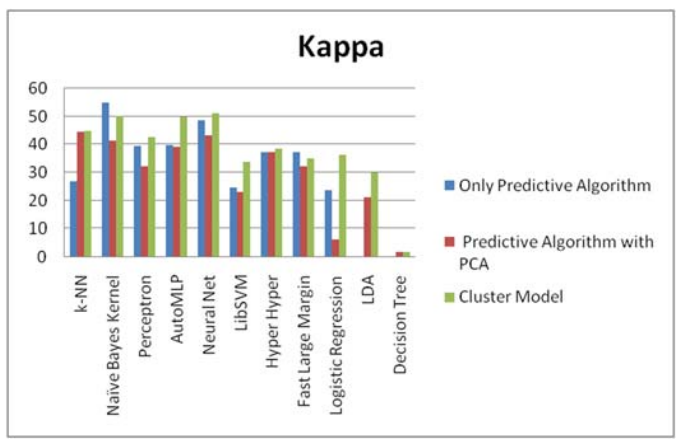

Fig. 2. Graph depicting Classifiers Performance with Kappa

Table II and Fig. 2 of Kappa Score clearly illustrate the improvement in Kappa score of each predictive algorithm when applied with Cluster Model. PCA on the other hand when applied before predictive algorithm does not improve results.

This can be further proved with Statistical t-test measures. Results reflect that the improvement is significant.

t-Test: Paired Two Sample for Means is conducted between the results obtained by Predictive Algorithms only and when the same algorithms were applied with PCA. Results are as follows:

t-Test: Paired Two Sample for Means

|  | Only Predictive Algorithm | Predictive Algorithm with PCA |
|---|---|---|
| Mean | 0.152090909 | 0.134181818 |
| Variance | 0.018099491 | 0.010727964 |
| Observations | 11 | 11 |
| Pearson Correlation | 0.597357168 |  |
| Hypothesized Mean Difference | 0 |  |
| df | 10 |  |
| t Stat | 0.538209984 |  |
| P(T<=t) one-tail | 0.301102379 |  |
| t Critical one-tail | 1.812461102 |  |
| P(T<=t) two-tail | 0.602204758 |  |
| t Critical two-tail | 2.228138842 |  |

Results depict that t Stat < t Critical two-tail

At 5% level of significance Ho 1.2 is accepted; hence there is no significant difference between Kappa values of predictive Algorithms when applied directly as compared to same predictive algorithms when applied using PCA There is also not much difference between variance values. PCA could not improve Kappa Score.

Further, t-Test: Paired Two Sample for Means is conducted between the results obtained by Predictive Algorithms with PCA and when the same algorithms were applied with cluster model. Results are as follows

t-Test: Paired Two Sample for Means

|  | Predictive Algorithm with PCA | Cluster Model |
|---|---|---|
| Mean | 0.13418182 | 0.203182 |
| Variance | 0.01072796 | 0.014775 |
| Observations | 11 | 11 |
| Pearson Correlation | 0.86725076 |  |
| Hypothesized Mean Difference | 0 |  |
| df | 10 |  |
| t Stat | -3.7797324 |  |
| P(T<=t) one-tail | 0.00180177 |  |
| t Critical one-tail | 1.8124611 |  |
| P(T<=t) two-tail | 0.00360354 |  |
| t Critical two-tail | 2.22813884 |  |

Results depict that t Stat > t Critical two-tail

At 5% level of significance Ho 1.2 is rejected; hence there is a significant difference between Kappa values of predictive Algorithms when applied using PCA at preprocessing as compared to same predictive algorithms when applied using Cluster Model at preprocessing. Cluster Model excels.

All the above results conclude that Predictive Algorithms when applied with cluster model improve its performance significantly.

## VI. CONCLUSION AND FUTURE SCOPE

The results prove that prediction performance for students' employability can be enhanced by applying Cluster Model, when dataset is large, unbalanced and multidimensional. Moreover, clustering applied on attributes set at pre-processing stage helps in parsimonious selection of variables and improves performance of predictive algorithms.

This paper also statistically analyzes and compares the results of predictive algorithms, when applied with proposed cluster model. The results clearly depicts that Cluster Model is superior to commonly used methods to predict students' employability and reducing dimensionality. Taking the base of proposed model more improvement can be made in future.